# Comment on: "Quantum interference of tunnel trajectories between states of different spin length in a dimeric molecular nanomagnet"

W. Wernsdorfer, Institut Néel - CNRS, 38042 Grenoble, Cedex 9, France

We recently submitted a correspondence to Nature Physics [1] with the intention to constructively correct a recently published paper [2]. This correspondence, written to be one-journal-page long, was obviously misunderstood [3] and we therefore feel that a more detailed correction has to be written. We show here that the data published in [2] are not consistent and unfortunately mostly wrong.

We will start with the main figure of the published letter, that is Fig. 3a of [2]. It shows the transverse field dependence of the tunnel splitting $\Delta_k$ of three tunnel resonances ($k$ = 0, 1S, and 1A, see [2]). It has been measured with a field sweep rate of 0.4 T/min, that is 0.0067 T/s [3]. The data for $k = 0$ are compared to our data in Fig. 1. We ask now the question whether these published values are reasonable. In order to check these values of $\Delta_k$, we use the Landau-Zener (LZ) equation, which gives the nonadiabatic LZ tunneling probability $P_k$ between two quantum states $m$ and $m'$ when sweeping the longitudinal field $H_z$ at a constant rate $dH_z/dt$ over an avoided energy level crossing:

$$P_k = 1 - \exp\left[-\frac{\pi \Delta_k^2}{2\hbar g \mu_B \Delta m \mu_0 dH_z/dt}\right] \quad (1)$$

where $\Delta m = |m - m'|$. We find that **any** value of Fig. 3a of [2] inserted into Eq. (1) yields $P_k$ values close to unity, using $\mu_0 dH_z/dt$ = 0.0067 T/s and $\Delta m$ = 14, 13, and 13 for $k$ = 0, 1S, and 1A, respectively. For example, for $H_x = 0$ and $\Delta_0 \approx 2*10^{-6}$ K, we find $P_k = 0.998$. This means that the tunnel probability of all resonances is so high that the spin system will relax always towards equilibrium, that is hysteresis loops should not be observed! This is obviously wrong because Fig. 1 of [2] shows clearly hysteresis. Therefore, all $\Delta_k$ values presented in Fig. 3a are inconsistent with the observation of hysteresis loops at the field sweep rates used by the authors of [2]. We recall that the Fig. 3a of [2] is the central measurement of the publication [2].

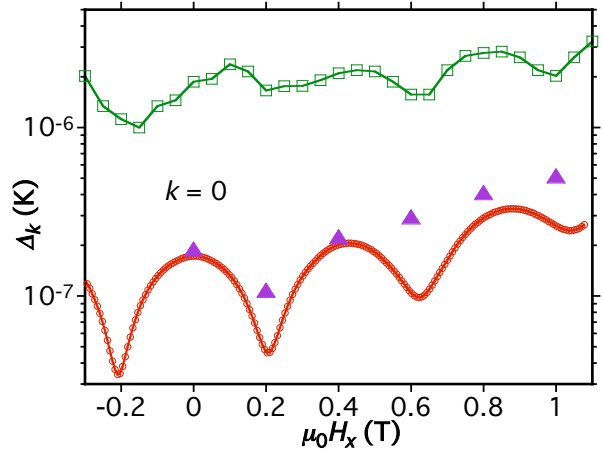

**Figure 1 Transverse field dependence of the tunnel splitting for the Mn$_{12}$ wheel.** Tunnel splittings $\Delta_k$ for $k = 0$ as a function of a transverse field applied along the hard axis of the Mn$_{12}$ wheel. The green open square-dots are the data reproduced from Fig. 3a in [2]. The red open dots are our data [1]. The violet triangles were determined from the data in Fig. 1e of [2], showing a wrong alignment of the transverse field because only a dip at 0.2 T is observed. Note that the difference between the data of [2] and our data is typically one order of magnitude. The influence of this difference on the Landau-Zener tunnel probability $P_{LZ}$ is huge because $\Delta_k$ is squared in an exponential, see Eq. 1.

Fig. 3c of [2] is a result of a numerical diagonalisation that should explain the results in Fig. 3a of [2]. Indeed, it suggests a rather good agreement between the model and the measurements! However, we have just shown that the $\Delta_k$ values of [2] are much too large to be in agreement with the observation of hysteresis loops. This means that the calculated values of $\Delta_k$ of Fig. 3c are also inconsistent with the observation of hysteresis loops and the rather good agreement between the model and the measurements is artificial.

The next question is whether Fig. 1e in [2] is consistent with quantum phase interference, which is the central effect discussed in the paper [2]. For that, we take the step height of the resonance at zero field ($k = 0$) and convert it into a tunnel splitting using Eq. (1). The result is plotted in Fig. 1 as a function of the transverse field (violet triangle-dots). Apart from a small dip at 0.2 T, a monotonous increase is observed, which is in contradiction with quantum phase interference, which was claimed to be observed. The same can be done with the other resonances leading to the same conclusion. Therefore, the field alignment in Fig. 1 cannot be correct.

In conclusion, all experimental results are not consistent with the claims of the paper. The reasons and other minor short-comings were explained in [1].

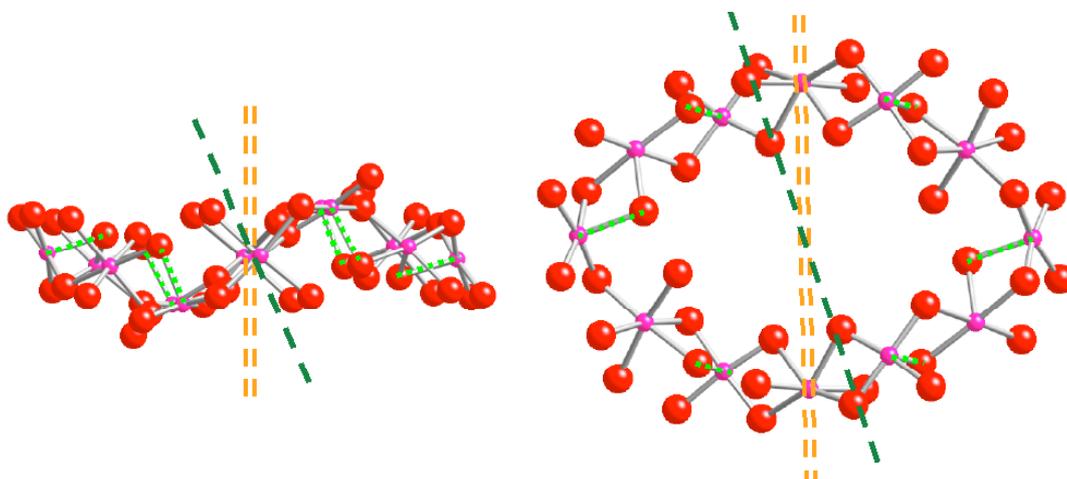

**Figure 2 Anisotropy axes of the $Mn_{12}$ wheel.** Illustration of the direction of the easy and hard anisotropy axes: **(left)** side view and **(right)** top view of the $Mn_{12}$ wheel. The light green dotted lines indicate the Jahn-Teller (JT) distortion axes, which indicate approximately the easy anisotropy axes of the $Mn^{3+}$ ions. The orange double-dotted line shows the directions, which were claimed to be the easy and hard magnetisation axis [2], respectively for the left and right figure. The dark green dashed lines show the easy and hard magnetisation axis that we proposed [1], which are given by the mean JT distortion axes.

We show now that the reply of the authors [3] contains many wrong statements. We focus only on the ***scientific*** part of the reply.

(1) The authors of [3] claim that we published a paper [4] saying the $Mn_{12}$ wheels should be treated as pairs of weakly coupled giant spins (dimers), with a two orders of magnitude difference between the inter- and intra-spin exchange coupling constants. In addition, the authors of [3] claim that this model was also confirmed in a subsequent work by Cano and collaborators [5]. Concerning our work, we would like to point out that the six independent exchange coupling constants $J$ obtained from DFT calculations on the $Mn_{12}$ wheel are all weak, spanning a range of only a few cm$^{-1}$ either side of zero. Given the inherent uncertainties of computed $J$ values, Ref. [4] did not conclude that the molecule can be considered as two weakly-coupled $Mn_6$ half-wheels. Concerning the work by Cano and collaborators [5], they find that all $J$ values are of comparable magnitude (see Table 1 of [5] - second column from the end. $J_6$ is the one that is supposed to be "two orders of magnitude" smaller, however, it is stronger than most of the other *J* values). Nevertheless, the authors of [3] claim that the work by Cano and collaborators [5] confirm the dimer model!

Concerning the possibility of Dzyaloshinskii-Moriya (DM) interaction, we repeat that [2] states that DM should be very small because the $Mn_{12}$ wheel has an inversion centre. However, DM interactions result from pairwise interactions of neighbouring spins, which do not have an inversion centre in the $Mn_{12}$ wheel. Therefore, we expect that DM interaction plays a role in the quantum dynamics. The authors of [2] argued that DM should come from defects.

Concerning our recent study of a $Mn_6$ single-molecule magnet [6], we did not use a dimer model but a model of 6 coupled spins.

Concerning our studies of the $[Mn_4]_2$ dimers [7], we used indeed a dimer model. However, in this case, the intramolecule couplings of the $Mn_4$-units are clearly two orders of magnitude larger than the coupling between the two $Mn_4$-units.

Concerning the $Mn_{12}$ wheel sample that we used, there is no doubt that we measured the same compound. In fact, we followed exactly the published method for $Mn_{12}$ wheels to obtain the sample and used standard methods to check the structure. In addition, we can reproduce exactly the same hysteresis loops when applying the field in the plane of the large face of the crystal, which was probably done to obtain Fig. 1c and 1e of [2]. We would like to emphasize that molecular magnets can be exactly reproduced when the synthetic method is published. Therefore, such samples can be studied independently by many groups. In particular, it allows a direct comparison of results.

Concerning the typographical error associated with the magnitude of the DM term, we agree with the authors that their value of 1.5 degrees is wrong in [2]. However, the new value of 0.036 degrees (i.e. *J* sin φ = 0.25 mK) is even worse. We are going to publish the correct values in the near future.

Concerning the claim of the authors [3] that we used the LZ formalism outside of its range of applicability in our original paper on the subject measured 10 years ago [8], we have to say that this is an unjustified claim. Indeed, these data were measured at field sweep rate about two orders of magnitude higher than those accessible by the authors of [2]. Therefore, the data for $k = n = 0$ and $n = 1$ are insight the LZ regime, and $n = 2$ was very close to it.

Concerning the following statement: *"We fully agree with Wernsdorfer's statement concerning the accuracy of the LZ method in terms of measuring tunnel splittings (Δ). Indeed, we add that molecules with larger Δ relax first. Consequently, the LZ formalism really only provides information concerning a subset of the molecules, i.e. those having the largest Δ. These facts are well known in the community, in part due to contributions from several authors of our Letter."* We would like to emphasize that the applicability rules of the LZ method, which we recalled in [1], should be followed even for "perfect" samples. The response of the authors in [3] addresses a different problem, namely that of tunnel splitting distributions, which are present in badly treated samples or samples with intrinsic disorder. If such effects were dominant in the $Mn_{12}$ wheels, it would be difficult to see quantum phase interference up to 1.4 T.

**References:**
[1] Wernsdorfer W., Correspondence on: Quantum interference of tunnel trajectories between states of different spin length in a dimeric molecular nanomagnet, http://arxiv.org/abs/0804.1246

[2] Ramsey, C. M., del Barco, E., Hill, S, Shah, S. J., Beedle, C. C., & Hendrickson, D. N. Quantum interference of tunnel trajectories between states of different spin length in a dimeric molecular nanomagnet, *Nature Phys.* **4**, 277-281 (2008).

[3] Ramsey, C. M., del Barco, E., Hill, S, Shah, S. J., Beedle, C. C., & Hendrickson, D. N., Reply to Wernsdorfer's post: "Correspondence on: Quantum interference of tunnel trajectories between states of different spin lenght in a dimeric molecular nanomagnet http://arxiv.org/abs/0806.1922

[4] Foguet-Albiol D, O'Brien TA, Wernsdorfer W, Moulton B, Zaworotko MJ, Abboud KA, Christou G, DFT computational rationalization of an unusual spin ground state in an Mn-12 single-molecule magnet with a low-symmetry loop structure. *Angew. Chem. Int. Ed.* **44**, 897-901 (2005).

[5] Cano, J., Costa, R., Alvarez, S., & Ruiz, E., Theoretical Study of the Magnetic Properties of an $Mn_{12}$ Single-Molecule Magnet with a Loop Structure: The Role of the Next-Nearest Neighbour Interactions. *J. Chem. Theory Comput.* **3**, 782-788 (2007)

[6] Carretta, T., et al., Breakdown of the Giant Spin Model in the Magnetic Relaxation of the $Mn_6$ Nanomagents, *Phys. Rev. Lett.* **100**, 157203 (2008)

[7] Wernsdorfer, W., Aliaga-Alcade, N., Hendrickson, D. N. & Christou, G. Exchange-biased quantum tunneling in a supramolecular dimer of single-molecule magnets. *Nature* **416**, 406–409 (2002) ; Tiron, R., Wernsdorfer, W., Foguet-Albiol, D., Aliaga-Alcalde, N., & Christou, G., Spin Quantum Tunneling via Entangled States in a Dimer of Exchange-Coupled Single-Molecule Magnets, *Phys. Rev. Lett.* **91**, 227203 (2003).

[8] Wernsdorfer, W. & Sessoli, R. Quantum phase interference and parity effects in magnetic molecular clusters. *Science* **284**, 133-135 (1999).

*Annex: For completeness, we reproduce here the second version of*
*http://arxiv.org/abs/0804.1246*

## Correspondence on: "Quantum interference of tunnel trajectories between states of different spin length in a dimeric molecular nanomagnet"
*[A shorted version of this correspondence was submitted to*
*Nature Physics the 31th March 2008]*

**W. Wernsdorfer,** Institut Néel - CNRS, 38042 Grenoble, Cedex 9, France

*To the Editor* — Ramsey et al.[1] report the observation of quantum interference associated with tunnelling trajectories between states of different total spin length in a dimeric molecular nanomagnet. They argue that the interference is a consequence of the unique characteristics of a molecular $Mn_{12}$ wheel, which behaves as a molecular dimer with weak ferromagnetic exchange coupling. The primary focus of this work is to shed light on the origin of a resonance, called $k = 1(A)$ in Ref. 1, which cannot be explained with the single-spin model. It is claimed to result from the avoided level crossing of the symmetric $|S = 7, M_S = −7\rangle$ and antisymmetric $|S = 6, M_S = 6\rangle$ states. Here we show that the Landau–Zener (LZ) formula, which links the tunnel probability with the tunnel splitting, can only be applied in a well-defined experimental region, which lays outside the region accessed by Ramsey and colleagues for the resonance $k = 1(A)$. Only a lower-limit estimate of the tunnel splitting can be obtained, showing that the observed transition cannot be explained with the dimer model.

We also present other shortcomings of the paper questioning the dimer model, and that the alignment of the magnetic field is crucial for observing quantum interference.

We recall that the nonadiabatic LZ tunneling probability $P_k$ between two quantum states $m$ and $m'$ when sweeping the longitudinal field $H_z$ at a constant rate $dH_z/dt$ over an avoided energy level crossing is given by

$$P_k = 1 - \exp\left[-\frac{\pi \Delta_k^2}{2\hbar g \mu_B \Delta m \mu_0 dH_z/dt}\right] \quad (1)$$

where $\Delta m = |m - m'|$. In order to apply quantitatively the LZ formula (Eq. 1), one has to check first the predicted field sweep rate dependence of the tunneling rate[2,3,4,5,6]. To do so, we first placed a crystal of the $Mn_{12}$ wheel in a high negative field $H_z$ to saturate the magnetization at 40 mK. We then swept the applied field at a constant rate over one of the resonance transitions and measured the variation of magnetization using a micro-SQUID (Fig. 1a-c). The fraction of molecules that reversed their spin was deduced from the step height, giving $P_k$. In order to test the LZ Eq. 1, we present the $\Delta_k$ values in Fig. 1d as a function of $dH_z/dt$. The LZ method is only applicable in the region of high sweep rates where $\Delta_k$ is independent of the field sweep rate. For the $k = 0$ resonance, this region is achieved for about $\mu_0 dH_z/dt > 0.1$ T/s. However, for the $k = 1(S)$ and $1(A)$ resonances, this region is not reached, even at $\mu_0 dH_z/dt = 1$ T/s (the measurements in Ref. 1 were performed at about 0.003 T/s). The deviations at lower sweep rates have been studied in detail[3,4,5,6] and are mainly due to reshuffling of internal fields[7,8]. Note that $\Delta_k$ obtained at lower sweep rates always underestimates the real $\Delta_k$ - it can therefore be used only as a lower-limit estimation.

Fig. 1e shows $\Delta_k$ as a function of a transverse field, approximately applied along the hard axis of magnetisation ($x$ axis) and measured at $\mu_0 dH_z/dt = 0.28$ T/s. The observed oscillations can be explained by quantum phase interference of two tunnel paths[9] and has been observed in other SMMs[2,10,11,12]. At $\mu_0 dH_z/dt = 0.28$ T/s the LZ method is applicable only for $k = 0$, that is for $k = 1(S)$ and $1(A)$ the obtained $\Delta_k$ values are lower-limits of the true $\Delta_k$.

Note that the values of $\Delta_k$ at $k = 0$ in Fig. 1e are one order of magnitude smaller than the values reported in Ref. 1. This is probably due to an error when applying Eq. 1 (see the triangles in Fig. 1e, which were determined form the data in Fig. 1e of Ref. 1).

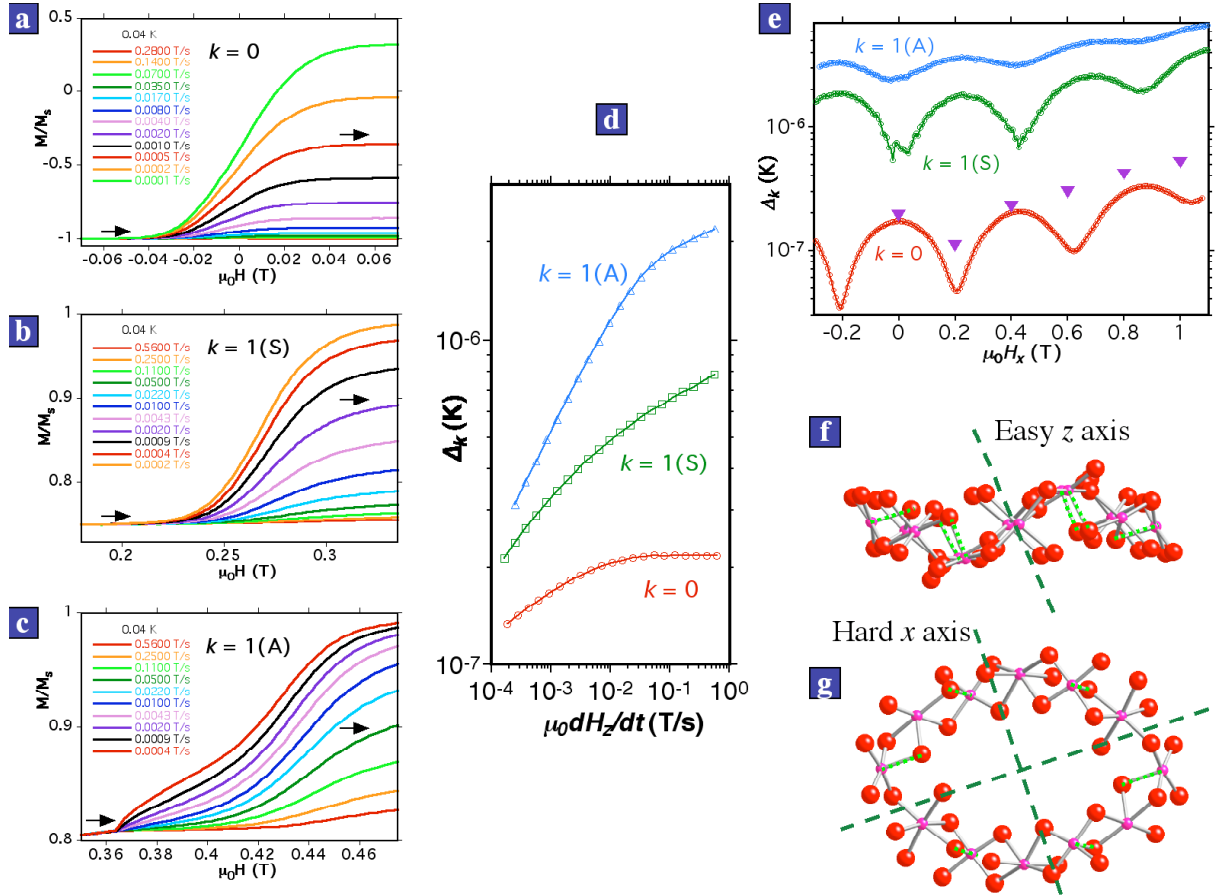

**Figure 1** Resonant tunnel, application of the LZ equation, quantum phase interference, and anisotropy axes of the $Mn_{12}$ wheel. **a – c,** Variation of magnetization for the indicated resonance transitions measured at several field sweep rates. In order to avoid a thermal run-away for $k = 1(S)$ and $1(A)$, we used the minor loop method described in Ref. 6. **d,** Field sweep rate dependence of the tunnel splitting $\Delta_k$ obtained from the LZ Eq. 1. The LZ method works in the region of high sweep rates where $\Delta_k$ is sweep rate independent. **e,** Tunnel splittings as a function of a transverse field applied along the hard axis of the $Mn_{12}$ wheel. Note that $\Delta_k$ for $k = 1(S)$ and $1(A)$ is only a lower-limit of the real value. The triangles were determined form the data in Fig. 1e of Ref. 1, showing a wrong alignment of the transverse field. **f – g,** Illustration of the direction of the hard and easy anisotropy axes. **f**, side view and **g**, top view of the $Mn_{12}$ wheel. The light green dotted lines indicate the Jahn-Teller (JT) distortion axes, which indicate the easy anisotropy axes of the $Mn^{3+}$ ions. The dark green dashed lines show the mean JT axis directions.

As noted in Ref. 1, the single spin model fails to explain the $k = 1(A)$ resonance. However, the dimer model of Ref. 1 also fails because it predicts $\Delta_k$ values of the $k = 0$ resonance which are one order of magnitude larger than our values in Fig. 1e. In addition, the $k = 1(A)$ resonance is forbidden in the dimer model. Ref. 1 proposes to explain the observed $k = 1(A)$ resonance by a Dzyaloshinskii-Moriya (DM) interaction. Ref. 1 states that DM should be very small because the $Mn_{12}$ wheel has an inversion centre. However, DM interactions result from pair vice interactions of neighbouring spins, which do not have an inversion centre in the $Mn_{12}$ wheel.

We would like to mention that the authors state that the easy axis of magnetization is along the axis of the $Mn_{12}$ wheel. However, the easy axis is given by the $Mn^{3+}$ Jahn-Teller (JT) distortion axes, which are tilted from the wheel axis by about 20° (Fig. 1f). Similarly, the hard axis of magnetisation was wrongly indicated in Ref. 1. This is important because quantum phase interference is very sensitive to the angle of the applied fields and a misalignment of the transverse and longitudinal fields by a few degrees would completely change the quantum phase interference.

Finally, Ref. 1 states that estimates for the strengths of the exchange interactions carried out for a similar compound[13] reveal that the exchange interaction between two manganese pairs at opposite sides of the wheel is over "two orders of magnitude" weaker than that of the other pairs of the wheel, splitting the molecule into two half-wheels of spin $S = 7/2$ each. We would like to point out that the six independent exchange coupling constants $J$ obtained from DFT calculations on the $Mn_{12}$ wheel are all weak, spanning a range of only a few cm$^{-1}$ either side of zero. Given the inherent uncertainties of computed $J$ values, Ref. 13 did not conclude that the molecule can be considered as two weakly-coupled $Mn_6$ half-wheels, which further questions the applicability of the dimer model in Ref 1.

**References:**

[1] Ramsey, C. M., del Barco, E., Hill, S, Shah, S. J., Beedle, C. C., & Hendrickson, D. N. Quantum interference of tunnel trajectories between states of different spin length in a dimeric molecular nanomagnet, *Nature Phys.* **4**, 277-281 (2008).

[2] Wernsdorfer,W. & Sessoli, R. Quantum phase interference and parity effects in magnetic molecular clusters. *Science* **284**, 133–135 (1999).

[3] Wernsdorfer W, Sessoli R, Caneschi A, Gatteschi D, Cornia A, Nonadiabatic Landau-Zener tunneling in Fe-8 molecular nanomagnets, *EuroPhys. Lett.* **50**, 552 (2000).

[4] Wernsdorfer W, et al., Landau-Zener method to study quantum phase interference of Fe-8 molecular nanomagnets. *J. Appl. Phys.* **87**, 5481-5486 (2000).

[5] Wernsdorfer W, Bhaduri S, Tiron R, Hendrickson DN, Christou G, Spin-spin cross relaxation in single-molecule magnets, *Phys. Rev. Lett.* **89**, 197201 (2002).

[6] Wernsdorfer W, Bhaduri S, Vinslava A, Christou G, Landau-Zener tunneling in the presence of weak intermolecular interactions in a crystal of Mn-4 single-molecule magnets, *Phys. Rev. B* **72**, 214429 (2005).

[7] W. Wernsdorfer, T. Ohm, C. Sangregorio, R. Roberta, D. Mailly, C. Paulsen, Observation of the distribution of molecular spin states by resonant quantum tunneling of the magnetization, *Phys. Rev. Lett.* **82**, 3903 (1999).

[8] Jie Liu, Biao Wu, Libin Fu, Roberto B. Diener, and Qian Niu, Quantum step heights in hysteresis loops of molecular magnets, *Phys. Rev. B* **65**, 224401 (2002).

[9] Garg, A. Topologically quenched tunnel splitting in spin systems without Kramers' degeneracy. *Europhys. Lett.* **22**, 205–210 (1993).

[10] Wernsdorfer, W., Soler, M., Christou, G. & Hendrickson D.N., Quantum phase interference (Berry phase) in single-molecule magnets of [Mn-12](2-). *J. Appl. Phys.* **91**, 7164-7166 (2002).

[11] W. Wernsdorfer, N. E. Chakov, and G. Christou, Quantum phase interference and spin-parity in Mn-12 single-molecule magnets, *Phys. Rev. Lett.* **95**, 037203 (2005).

[12] Lecren L, Wernsdorfer W, Li YG, Roubeau O, Miyasaka H, Clerac R, Quantum tunneling and quantum phase interference in a [(Mn2Mn2III)-Mn-II] single-molecule magnet, *J. Am. Chem. Soc.* **127**, 11311-11317 (2005).

[13] Foguet-Albiol D, O'Brien TA, Wernsdorfer W, Moulton B, Zaworotko MJ, Abboud KA, Christou G, DFT computational rationalization of an unusual spin ground state in an Mn-12 single-molecule magnet with a low-symmetry loop structure. *Angew. Chem. Int. Ed.* **44**, 897–901 (2005).